\documentclass[12pt]{article}

\usepackage{cite}
\usepackage[english]{babel}
\usepackage{graphicx,rotating}
\newcommand {\stl}  {\sin^2 \theta_{\rm eff}^l}
\newcommand{\MZ}      {m_{\mathrm{Z}}}
\newcommand{\Afbzb}     {A^{0,\,{\rm b}}_{\rm {FB}}}
\newcommand{\Afbzq}     {A^{0,\,{\rm q}}_{\rm {FB}}}
\newcommand{\Afbb}    {A_{\mathrm{FB}}^{\rm b}}
\newcommand{\Afbc}    {A_{\mathrm{FB}}^{\rm c}}
\newcommand{\cA} {{\cal A}}
\def\GeV{\ifmmode {\,\mathrm{ Ge\kern -0.1em V}}\else
                   \textrm{Ge\kern -0.1em V}\fi}%
\begin{document}



\begin{titlepage}

\def\thefootnote{\fnsymbol{footnote}}
\pagenumbering{arabic}
\begin{flushright}
23rd November 2004\\[1ex]
FERMILAB--Pub--04/352--T\\
DESY 04--225\\
\end{flushright}
\vspace*{2.cm}
\begin{center}
\Large 
\boldmath
{\bf
Corrections to Quark Asymmetries at LEP
} \\
\unboldmath
\vspace*{2.cm}
\normalsize { 
{\large
{\sc
A.~Freitas$^{1}$%
\footnote{email: afreitas@fnal.gov},
K.~M\"onig$^{2}$%
\footnote{email: Klaus.Moenig@desy.de}
}
}

\vspace*{1cm}

{\sl
$^1$ Fermi National Accelerator Laboratory, Batavia, IL 60510-500, USA

\vspace*{0.4cm}

$^2$ DESY, Platanenallee 6,  D--15738 Zeuthen, Germany
}
}
\end{center}
\vspace{\fill}
\begin{abstract}
\noindent
The most precise measurement of the weak mixing angle $\stl$ at LEP is from
the forward-backward asymmetry $e^+ e^- \rightarrow b \overline{b}$ at the
Z-pole. In this note the QED and electroweak radiative corrections to obtain
the pole asymmetry from the measured asymmetry for b- and c-quarks
have been calculated using {\tt ZFITTER}, which has been amended to allow a 
consistent treatment of partial two-loop corrections for the b-quark final 
asymmetries.

A total correction of $\delta \Afbb =  0.0019 \pm 0.0002$ and
$\delta \Afbc =  0.0064 \pm 0.0001$ has been found, where the remaining
theoretical uncertainty is much too small to explain the apparent discrepancy
between $\stl$ obtained from $\Afbb$ and from the left-right asymmetry at SLD.
\end{abstract}
\vspace{\fill}

\vspace{\fill}
\end{titlepage}
%
%
%
\def\thefootnote{\arabic{footnote}}
\setcounter{footnote}{0}
\setcounter{page}{1}

\section{Introduction}
At LEP and SLD the effective electroweak mixing angle at the Z-scale, $\stl$,
can be measured using several different asymmetries \cite{lepewep}. The two
most precise measurements of this quantity are obtained from the the
left-right asymmetry with a polarised electron beam at SLD and the
forward-backward asymmetry for b-quarks at LEP. Both measurements provide a
relative precision on $\stl$ of around $10^{-3}$. This is significantly more
precise than the expected loop effects which allows for example the estimation
of the Higgs boson mass from the electroweak precision data. On the other hand
this very high precision also requires a good understanding of all higher
order corrections like photon radiation, photon exchange, mass effects etc.

In this note the procedure to correct the measured forward-backward 
asymmetries for b-quarks will be described with special attention to the 
recent modifications.

\section{Correction Procedure}
In the electroweak fits the experimental measurements are not used
directly but instead so called pseudo-observables are used that are obtained
from the measurements with some almost model independent corrections.
For the b-asymmetry this pseudo-observable is the pole asymmetry 
$\Afbzb$. This pole asymmetry can be viewed as the b-asymmetry on the Z-peak
without photon exchange, QED and QCD corrections and taking only the real
parts of the Z-fermion couplings. The weak mixing angle is then given in terms
of the pole asymmetry as 

\begin{eqnarray}
\Afbzb & = & \frac{3}{4} \cA_e \cA_b, \label{eq:afb}\\
\cA_f & = & \frac{2g_{Vf}g_{Af}}{g_{Vf}^2+g_{Af}^2}, \nonumber \\
\frac{g_{Vf}}{g_{Af}} & = & 1-4 Q_f \sin^2 \theta_{\rm eff}^f, \nonumber
\end{eqnarray}
where $g_V$ and $g_A$ are the effective coupling constants
of the weak neutral current.

The QCD corrections \cite{ref:QCDcor,ref:QCDupd} 
arise mainly from the smearing of the
event axis due to gluon radiation. Their size thus depends strongly on the
experimental selection procedure. For this reason the QCD corrections are
already performed by the experiments and corrected values are provided for
combination. The procedure is described in detail in \cite{ref:QCDcor}.

The energy dependence of the asymmetries is given by the contribution from
the $\gamma$--Z interference. It is numerically large, but can be
predicted with negligible theoretical uncertainty. 
In the combination procedure all experimental asymmetries are first corrected
to a centre of mass energy of $ \sqrt {s} = 91.26\,\GeV$ assuming a Standard
Model energy dependence as predicted by {\tt ZFITTER} \cite{ref:zfitter}.
In a last step the LEP combined value of $\Afbb (91.26 \GeV)$ is then
corrected to the pole asymmetry $\Afbzb$ \cite{NIM}. 
For c-quarks exactly the same procedure is followed.

This correction is again done using {\tt ZFITTER}. This program allows to 
calculate realistic observables and  as well as pseudo-observables. The total
correction is calculated as 
\[
\delta \Afbb = \left( \Afbzb - \Afbb \right)_{\rm ZFITTER}
\]
so that the measured value of $\Afbzb$ can be expressed as
\[
\Afbzb( {\rm meas}) = \Afbb ({\rm meas}) + \delta \Afbb.
\]

For clarity the correction is split into three parts:
\begin{enumerate}
\item energy shift from $ \sqrt {s} = 91.26\,\GeV$ to $ \sqrt {s} =
  \MZ$;
\item QED corrections;
\item other corrections including $\gamma$-exchange and 
  $\gamma-{\rm Z}$-interference, mass effects and imaginary parts of
  couplings. 
\end{enumerate}
The QED corrections affect the asymmetries mainly by the change in
centre of mass energy due to the initial state radiation. Since the
average running energy of LEP was slightly above the Z-mass the QED
corrections and the energy correction are partially cancelling. 
For these corrections it is also clear that they should be treated as
an additive correction. With the small experimental error on the b
asymmetry it makes, however, numerically no difference if the correction is
treated as additive or multiplicative.

\section{{\tt ZFITTER} Modifications}
\label{sec:zfitter}

Recently, complete electroweak two-loop results for the prediction of the
$W$-boson mass, $M_{\rm W}$ \cite{mwf,mwb}, and exact fermionic results for the
two-loop corrections to the effective leptonic weak mixing, $\stl$,
\cite{sineff} became available. Here the fermionic two-loop
corrections denote all two-loop contributions with at least one closed fermion
loop. These results improve on the prediction for the precision
pseudo-observables in the Standard Model with respect to the previously known
partial results for electroweak two-loop corrections using an expansion for
large values of the top-quark mass up to next-to-leading order \cite{mtexp}.
These latter results had been incorporated into {\tt ZFITTER} from version 5.10
upwards \cite{ref:zfitter}. Complete two-loop results for the $Z$-boson partial
widths are still missing.

The fermionic two-loop corrections to $M_{\rm W}$ \cite{mwf} were implemented
into {\tt ZFITTER} in the version 6.36. The version 6.40 incorporates the
complete two-loop corrections to $M_{\rm W}$ \cite{mwb} including new partial
three-loop corrections of order ${\cal O}(\alpha^3)$ and ${\cal O}(\alpha^2
\alpha_{\rm s})$ \cite{faisst}. In the same version also the fermionic two-loop
corrections to the pseudo-observable $\stl$
\cite{sineff} are implemented. Internally, the pseudo-observables are
computed in the subroutine {\tt ZWRATE} in the package {\tt DIZET}.

The interfaces {\tt ZUTHSM}, {\tt ZUTPSM}, {\tt ZULRSM} and {\tt ZUATSM}, on
the other hand, calculate the cross-sections and asymmetries directly from
the Standard Model predictions for the weak vertex form factor computed in the
subroutine {\tt ROKANC} \cite{ref:zfitter}. It is important to observe that the
weak form factors, denoted $\rho_{ef}(s,t)$, $\kappa_e(s,t)$, $\kappa_f(s,t)$
and $\kappa_{ef}(s,t)$, depend on the final state fermion type $f$. The weak
corrections for the $b\bar{b}$ final state are substantially different than for
the other flavours, since the vertex loop corrections for the $Zb\bar{b}$
vertex involve heavy internal top-quark propagators. This peculiarity is
consistently treated in {\tt ZFITTER} up to one-loop order. Up to now, however,
no two-loop results for the electroweak corrections to the $Zb\bar{b}$ vertex
are available. This is already true for the previously known leading-$m_{\rm
t}$ corrections \cite{mtexp}. In this case, {\it i.e.} for ${\tt INDF}=9$, all
versions of {\tt ZFITTER} up to 6.40 calculate {\it all} four form factors in
{\tt ROKANC} in one-loop approximation.

While this is the best possible treatment for the $Zb\bar{b}$ vertex that we
can achieve today, 
it produces inconsistencies for the initial $Ze^+e^-$ form factors
when including electroweak two-loop corrections, {\it i.e.} for ${\tt AMT4} \ge
4$. The reason is that in {\tt ROKANC} the $Ze^+e^-$ from factors for all other
final states will be generated including two-loop corrections, while for the
$b\bar{b}$ final state only one-loop corrections are used.

This mismatch also affects the {\tt ZFITTER} interfaces {\tt ZUXSA}, {\tt
ZUTAU} and {\tt ZUXSA2}, which use the language of effective couplings
\cite{ref:zfitter}, since they are defined to coincide exactly with the
complete Standard Model prediction in {\tt ROKANC} if the effective couplings
coincide with their Standard Model analogue.

The problem has been alleviated in the newest version {\tt ZFITTER} 6.41
\cite{zfit641}. In
contrast the older implementations, $\kappa_e(s,t)$ and $\kappa_f(s,t)$ are
not treated symmetrically anymore for ${\tt INDF}=9$, but two-loop electroweak
corrections are included in $\kappa_e(s,t)$ for ${\tt AMT4} \ge 4$, yet not in
$\kappa_b(s,t)$. The treatment of $\rho_{ef}(s,t)$ and $\kappa_{ef}(s,t)$ has
been changed accordingly. Here one can use the fact that the presently known
two-loop contributions factorise into initial-state and final-state
corrections. The changes in the code for {\tt ROKANC} affect both the treatment
of the previously available leading-$m_{\rm t}$ corrections for 
${\tt AMT4} =4$, as well as the new corrections for $\stl$.
Numerically these modifications lead to an upward shift of about 0.0006 
for the prediction of $\Afbb$ compared to previous {\tt ZFITTER} versions.

The new two-loop corrections \cite{mwf,mwb,sineff}, which do not rely on a
large-mass expansion, can be accessed through the flag {\tt AMT4}. The setting
${\tt AMT4}=5$ corresponds to the status of the version {\tt ZFITTER} 6.36,
which includes the complete fermionic corrections to the $W$ mass \cite{mwf}
and has been used for the summer 2001 LEP electroweak fits \cite{lep2001}. The
assignment ${\tt AMT4}=6$, used from summer 2004 onwards,
enables the calculation of $M_{\rm W}$ including
complete two-loop and leading three-loop corrections \cite{mwb,faisst} and the
inclusion of the new fermionic two-loop corrections to the effective weak
mixing angle. The estimated theoretical uncertainties for these two quantities
can be simulated by varying the flags ${\tt DMWW}$ (for $M_{\rm W}$ and ${\tt
  AMT4}=5,6$) and ${\tt DSWW}$ (for $\stl$ and
only ${\tt AMT4}=6$) between $-1$ and 1, respectively.

\section{Results}

The corrections to the quark asymmetries are summarised in Table
\ref{tab:asycor}. For comparison also the corrections for s-quarks are given.
For these values the full two-loop corrections on $\stl$ ({\tt AMT4=6}) are
used. In this case it is assumed that the couplings of the $Zf\bar{f}$ vertex
factorise between the initial and the final state. The flavour specific
corrections for the $Zb\bar{b}$ vertex are not yet calculated, however they
are highly suppressed because of the small b-quark charge and because $\cA_b$
is so close to 1 (see eq. \ref{eq:afb}).  For c- and s-quarks no
approximations are involved.  Up to now corrections of $\delta \Afbb = 0.0025$
and $\delta \Afbc = 0.0062$ have been used \cite{lepewep}. While there is no
significant change for $\Afbc$, for the b-asymmetry the total correction is
0.0006 below that value so that the LEP-combined result of $\Afbzb$ will
decrease by that amount.
\begin{table}[bht]
\begin{center}
\begin{tabular}{|l|l|l|l|}
\hline
Source                  & $\delta \Afbc$
         & $\delta \Afbb$ & $\delta {A_{\mathrm{FB}}^s}$\\
\hline
$\sqrt{s} = \MZ $       & $ -0.0035$  & $ -0.0014 $  & $ -0.0014 $  \\
QED corrections         & $ +0.0107$  & $ +0.0039 $  & $ +0.0038 $  \\
other                   & $ -0.0008$  & $ -0.0006 $  & $ -0.0003 $  \\
\hline
Total                   & $ +0.0064$  & $ +0.0019 $  & $ +0.0021 $  \\
\hline
\end{tabular}
\end{center}
\caption[Corrections to be applied to the quark asymmetries.]
{Corrections to be applied to the quark asymmetries as 
   $\Afbzq = A_{\mathrm{FB}}^{\rm q}({\rm pk})
  + \delta A_{\mathrm{FB}}$.
  ``other'' denotes corrections due to $\gamma$ exchange, $\gamma-$Z
   interference, quark-mass effects and imaginary parts of the couplings.
}
\label{tab:asycor}
\end{table}

To verify that these corrections are reliable several cross checks
have been made. If the Higgs mass is varied between 100\,GeV and 1\,TeV
the corrections stay constant. The same is true if the top-mass and
$\alpha(\MZ)$ are varied within several standard deviations. 
Also when instead of
the full 2-loop corrections only the leading corrections by Degrassi
{\it et al.} \cite{mtexp} are used, that are implemented in {\tt ZFITTER} since
long ({\tt AMT4=4}) none of the values in Table \ref{tab:asycor} change.

If one uses only the full 1-loop corrections to $\stl$ ({\tt AMT4=3}) 
the b-quark treatment in {\tt ZFITTER} is exact. 
In this case the total correction increases
by 0.00015. However the difference between the b-quark and the s-quark
correction stays constant showing that this is a genuine 2-loop effect
and not an artefact of the involved approximations.

With the public version of {\tt TOPAZ0} \cite{topaz} it is not possible to
reproduce Table \ref{tab:asycor}, since final state QCD corrections
cannot be switched off. However the initial and final state
deconvoluted asymmetries as well as $\Afbzb$ using {\tt TOPAZ0} are given in
\cite{prcalc} (Tables 6 and 14). From these values the critical
correction labelled ``other'' in Table \ref{tab:asycor} can be calculated to
be $-0.0005$, well in agreement with the {\tt ZFITTER} value.

\subsection{Systematic uncertainties}
To assess the systematic uncertainty from QED corrections the relevant flags
in {\tt ZFITTER} have been varied. The only flag for which a variation of the
result has been found was {\tt FBHO}, which describes the treatment of fermion
pair radiation in the asymmetry calculation. From this an error of

\begin{eqnarray*}
\Delta (\delta \Afbb)({\rm QED}) & = & 0.00017, \\
\Delta (\delta \Afbc)({\rm QED}) & = & 0.00011,
\end{eqnarray*}
has been derived. Especially no uncertainty from the choice of the radiator
function (flag {\tt FOT2}) has been found.

For the b-quark the flavour specific two-loop corrections have not been
calculated yet. As an estimate of the uncertainty from these diagrams the
difference between the total correction for b- and s-quarks has been used
leading to 
\begin{eqnarray*}
\Delta (\delta \Afbb)({\rm 2-loop\,b}) & = & 0.00016.
\end{eqnarray*}
To test the uncertainty due to the universal higher order corrections the
flags {\tt DMWW} and {\tt DSWW} mentioned in section \ref{sec:zfitter} have
been varied and no significant change in the asymmetry corrections has been 
seen.
This results in a total uncertainty of the QED and electroweak corrections to
the b- and c-quark asymmetry of
\begin{eqnarray*}
\Delta (\delta \Afbb) & = & 0.0002, \\
\Delta (\delta \Afbc) & = & 0.0001.
\end{eqnarray*}

\section{Conclusions}
After correcting some inconsistencies in the treatment of b-quarks in 
{\tt ZFITTER},
the QED and electroweak corrections to obtain the pole asymmetry from the
measured, QCD corrected, forward-backward asymmetry at the Z peak have been
calculated. 
Total corrections of
\begin{eqnarray*}
\delta \Afbb & = & 0.0019 \pm 0.0002, \\
\delta \Afbc & = & 0.0064 \pm 0.0001,
\end{eqnarray*}
have been found. The total corrections  are only slightly larger than the
experimental errors while their uncertainties are about an order of magnitude
smaller. It is thus inconceivable that these corrections can explain the
apparent difference in $\stl$ obtained from the left-right asymmetry at SLD and
from $\Afbb$.

\section*{Acknowledgements}
We would like to thank the LEP heavy flavour working group,
A.~Arbuzov, M.~Awramik, M.~Czakon, M.~Gr\"unewald, G.~Passarino,
S.~Riemann and T.~Riemann for useful discussions.


\begin{thebibliography}{10}
\bibitem{lepewep}
The LEP collaborations, {\it A Combination of Preliminary Electroweak 
Measurements and Constraints on the Standard Model},
CERN-EP/2003-091, hep-ex/0312023
%
\bibitem{ref:QCDcor}
LEP Heavy Flavour Working Group Collaboration, D. Abbaneo {\it et~al.},
Eur. Phys. J. {\bf C4}  (1998) 185.
%
\bibitem{ref:QCDupd}
The LEP Heavy Flavour Group, {\it Final input parameters for the LEP/SLD heavy
  flavour analyses,} LEPHF/01-01, \\
  http://www.cern.ch/LEPEWWG/heavy/lephf0101.ps.gz.
%
\bibitem{ref:zfitter}
D.~Y. Bardin {\it et~al.},
Comput. Phys. Com. {\bf 133}  (2001) 229.
%
\bibitem{NIM}
ALEPH, DELPHI, L3, OPAL Collaboration, 
Nucl. Instrum. Meth. {\bf A378}  (1996) 101.
%
\bibitem{mwf}
A.~Freitas, W.~Hollik, W.~Walter and G.~Weiglein,
Phys.\ Lett.\ B {\bf 495} (2000) 338
[Erratum-ibid.\ B {\bf 570} (2003) 260],
Nucl.\ Phys.\ B {\bf 632} (2002) 189
[Erratum-ibid.\ B {\bf 666} (2003) 305];\\
M.~Awramik and M.~Czakon,
Phys.\ Lett.\ B {\bf 568} (2003) 48.
%
\bibitem{mwb}
M.~Awramik and M.~Czakon,
Phys.\ Rev.\ Lett.\  {\bf 89} (2002) 241801;\\
A.~Onishchenko and O.~Veretin,
Phys.\ Lett.\ B {\bf 551} (2003) 111;\\
M.~Awramik, M.~Czakon, A.~Onishchenko and O.~Veretin,
Phys.\ Rev.\ D {\bf 68} (2003) 053004;\\
M.~Awramik, M.~Czakon, A.~Freitas and G.~Weiglein,
Phys.\ Rev.\ D {\bf 69} (2004) 053006.
%
\bibitem{sineff}
M.~Awramik, M.~Czakon, A.~Freitas and G.~Weiglein,
hep-ph/0407317,\\
see also
hep-ph/0408207,
hep-ph/0409142.
%
\bibitem{mtexp}
G.~Degrassi, P.~Gambino and A.~Vicini,
Phys.\ Lett.\ B {\bf 383} (1996) 219;\\
G.~Degrassi, P.~Gambino and A.~Sirlin,
Phys.\ Lett.\ B {\bf 394} (1997) 188;\\
G.~Degrassi and P.~Gambino,
Nucl.\ Phys.\ B {\bf 567} (2000) 3.
%
\bibitem{faisst}
M.~Faisst, J.~H.~K\"uhn, T.~Seidensticker and O.~Veretin,
Nucl.\ Phys.\ B {\bf 665} (2003) 649.
%
\bibitem{lep2001}
The LEP EWWG and SLD Heavy Flavour and Electroweak Groups, D.~Abbaneo {\it et
al.}, hep-ex/0112021.
%
\bibitem{zfit641}
{\tt /afs/cern.ch/user/b/bardindy/public/ZF6\_41.}
%
\bibitem{topaz}
G. Montagna {\it et~al.},
Comput. Phys. Com. {\bf 117} (1999) 278.
%
\bibitem{prcalc}
D.~Y.~Bardin, M.~Gr\"unewald and G.~Passarino,
hep-ph/9902452.
\end{thebibliography}
\end{document}